\documentclass[]{aa} 
\usepackage{epsfig}
\def\kms{km~s$^{-1}${}}

\thesaurus{03 (11.03.1; 11.03.4; 12.04.1; 13.25.2) }

\title{A multi-wavelength analysis of the cluster of galaxies ABCG 194 }

\author {E.~Nikogossyan\inst{1}, F.~Durret\inst{2,3}\and  
D.~Gerbal\inst{2,3} \and F.~Magnard\inst{2}}

\offprints{F. Durret, durret@iap.fr }
\institute{
Byurakan Observatory, Aragatsotn Province, 375433, Armenia
\and
Institut d'Astrophysique de Paris, CNRS, 98bis Bd Arago, F-75014 Paris, France
\and
DAEC, Observatoire de Paris, Universit\'e Paris VII, CNRS (UA 173),
F-92195 Meudon Cedex, France
}
\date{Received, 1999; accepted,}

\begin{document}
\maketitle

\begin{abstract}

We present a morphological and structural analysis of the Richness
zero cluster ABCG 194, known as a ``linear cluster''.  This study is
based on a catalogue of 97 galaxies with B magnitudes and redshifts
belonging to the cluster, a ROSAT PSPC image and radio data.

We show that the overall large scale structure is rather smooth and
comparable at optical and X-ray wavelengths. The cluster is elongated
along PA $\approx 50$; however it appears as ``linear'' when taking
into account only galaxies in the very central part (the axial ratio
varies from 0.2 in the central region to 0.8 for a larger region). We
have obtained the galaxy density profile and despite the very low
X-ray emission the X-ray emitting gas density profile.  We have
estimated the X-ray gas and dynamical masses up to the limiting radius
R$_{\mathrm{L}}$ of the PSPC image (respectively $9\ 10^{12}$ and 8 
10$^{13}$ M$_\odot$) and the stellar mass up to R$_{\mathrm{L}}$ and
to 3 Mpc (3 $10^{12}$ and 7.5 10$^{12}$ M$_\odot$).

At smaller scales, the analysis of both positions and velocities
reveals the existence of several groups but which are not strongly
linked dynamically; however a main structure with a nearly gaussian
velocity distribution is exhibited.  The velocity dispersion is small
($\sigma_{los} $ = 360 \kms). A wavelet analysis of the X-ray image
reveals no large scale substructures. Small scale X-ray sources are
detected, mainly corresponding to individual galaxies; we give an
estimate of their luminosities.  The lack of strong substructuring
suggests that ABCG 194 is overall a relaxed cluster.

ABCG 194 is a poor and cold cluster; we compare how its
characteristics fit into various correlations found in the literature,
but generally for richer/hotter clusters, such as the 
$\sigma$-T$_{\mathrm{X}}$, L$_{\mathrm{X}}$-T$_{\mathrm{X}}$, 
L$_{\mathrm{X}}$-$\sigma$ relations or the ratios of various kinds of masses.

\end{abstract}

\section{Introduction}\label{intro}

It is presently believed that large scale structures (in particular
clusters of galaxies) are formed through hierarchical clustering.
From this point of view, clusters are formed by merging of smaller
clusters or by groups falling onto a larger cluster. However, the
effect of merging is not the same if a cluster undergoes a major
merger or accretes small groups or even single galaxies 
(Salvador-Sol\'e et al. 1998).

Notice that merging does not occur isotropically around a cluster.
Groups tend to fall onto clusters along filaments, thus explaining the
preferential orientations often observed (see e.g. West et al. 1995,
Durret et al. 1998).  Therefore the true orientation and the true
ellipticity are interesting quantities to derive.  The imprints of
such merging events can be seen directly on the X-ray images, but also
in substructures which can be detected at optical (density maps,
velocity structure), X-ray and radio wavelengths; this was first
noticed by Baier (1977) from optical data.

Recent studies have revealed that about 50\% of clusters show
multi-component structure both in their galaxy distribution (Bird
1994, Escalera et al.  1994, West 1994, Maccagni et al.  1995) and in
X-rays (Mohr et al.  1993, Grebenev et al.  1995, West et al. 1995,
Zabludoff \& Zaritsky 1995). However one may notice that, when the
amount of data grows and techniques of analysis are improved, this
percentage appears to grow too (Girardi et al. 1997). This leads to
question the dynamical meaning of these substructures.

The properties of relatively poor and cool clusters are somewhat more
difficult to analyze because of their limited signal, but prove to be
interesting as a link between rich clusters and groups.  Their
relaxation time scales are larger and one can wonder if they are
sufficiently virialized (e.g. Mahdavi et al.  1999 and references
therein).  The properties of the X-ray emitting gas can be different
from those of rich/hot clusters (see e.g.  David et al.  1996).  In
particular, various relations between optical and X-ray properties (as
the $\sigma$ - T$_{\rm X}$ relation which seems linked to global
properties), X-ray/X-ray relations (as the L$_{\rm X}$-T$_{\rm X}$
relation) or X-ray to dynamical properties (as the X-ray gas fraction)
may differ from those of rich clusters (Mahdavi et al.  1997,
Markevitch 1998, Arnaud \& Evrard 1999): they are probes to test
cluster formation and evolution models, particularly the fine modeling
of the gas heating (Cavaliere et al.  1998).  Very few low
temperature clusters are included in these relations, so adding even a
single cluster at the border-line is important.

We present a multi-wavelength analysis of the cluster ABCG 194 based
on optical data taken from the literature and on X-ray data from the
ROSAT archive, coupled with different techniques of analysis.

ABCG 194 is a linear cluster of richness 0 and Bautz-Morgan type III
(Struble \& Rood 1982) with central coordinates $\alpha_{1950} = 1^h
23.4^{mn}$ and $\delta_{1950} = -1^\circ 36'$ (Chapman et al.  1988,
hereafter CGH).  Its average redshift is 0.018, corresponding to a
heliocentric velocity of 5340 km/s, and to a spatial scaling of 1.87
Mpc/degree (H$_{0}$= 50 km/s/Mpc, q$_{0}$= 0).  The velocity interval
corresponding to the cluster is 4000-6600 km/s according to CGH, but
we will show in section 4.2 that it is in fact 4300-6200 km/s.

We have somewhat arbitrarily divided our analysis into two scales: the
global aspect at large scale, and the smaller scale features. The
first description is related to the cluster as an astrophysical object
by itself, while for the second description, which takes into account
as much as possible the kinematics, dynamical processes are important.

\section{The data}

\begin{figure}[tbp]
\centerline{
\psfig{figure=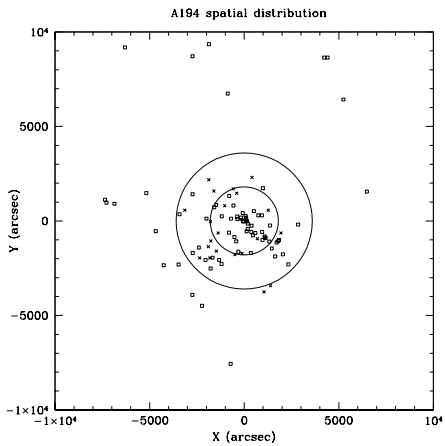,height=6cm,angle=0}}
\caption{Positions of the sample galaxies relative to the cluster center. 
Squares and crosses indicate galaxies from the CGH and BCG samples 
respectively. The circles have 30 arcmin and 1$^\circ$ radii.}
\protect\label{xya194}
\end{figure}

\subsection{Optical data}

Our optical analysis is based on the following data:

- Chapman et al. (1988) have observed a list of 74 cluster member
galaxies in ABCG 194 (i.e.  with velocities in the 4000-6600 km/s
range), located within a radius of 3$^\circ$ (5.6 Mpc) around the
cluster center.  The CGH sample is 97\% complete for galaxies brighter
than B=16.7 in a region of radius R=30 arcmin around the cluster
center, and 80\% complete for galaxies brighter than B=15.5 in the
entire region.

Out of the 74 galaxies in this sample, 48 and 62 are located within
radii of 30 arcmin and 1$^\circ$ of the cluster center
respectively. For all these galaxies, CGH give the following data:
coordinates, radial velocity, B magnitude, morphological type, major
axis position angle and ellipticity.

- We have added to this sample 22 optically fainter galaxies selected
from the CF2+SSRS2 redshift survey, for which Barton et al. (1998)
give positions and redshifts (hereafter the BCG sample); these authors
determined that the region of 1$^\circ$ radius around the ABCG 194
cluster is 100\% complete for galaxies brighter than
m$_{Zw}=16.46$. For the three galaxies with numbers 15, 30 and 54 in
the CGH list, we take the radial velocity from the Barton et
al. (1998) paper, where the error is smaller than in CGH. 

- The coordinates and redshift for an additional (Seyfert) galaxy were
taken from Knezek \& Bregman (1998). 

The total sample therefore includes 97 galaxies located within the
following range of positions relative to the cluster center: $-7360
\leq$X$\leq$6480 and $-7560 \leq$Y$\leq$8710 arcsec.

\subsection{X-ray data}

\begin{figure}[tbp]
\caption{Digital Sky Survey image superimposed with ROSAT X-ray image
contours (1, 2, 3 and 4 $\sigma$) superimposed.}
\protect\label{opt_x}
\end{figure}

\begin{figure}[tbp]
\caption{Digital Sky Survey image superimposed with NVSS radio (1.4 GHz) image
contours (from 12 $\sigma$ up by a factor of 2).}
\protect\label{opt_radio}
\end{figure}

Our X-ray study is based on a ROSAT PSPC image of ABCG 194 retrieved
from the ROSAT data bank; the exposure time was 24482 seconds
(P.I. Murray \& Stephens). This image was processed with the Snowden
software (Snowden et al. 1994). Obviously, this cluster is not a
strong X-ray emitter, since in spite of the relatively high exposure
time the total amount of counts is 8530, which is small. We show in
Fig.  \ref{opt_x} a superposition of the optical and X-ray maps, and
in Fig.  \ref{opt_radio} that of the optical and radio maps. Notice
that there is a zoom by a factor of $\sim$3 between these two images.

\begin{figure}
\centerline{\psfig{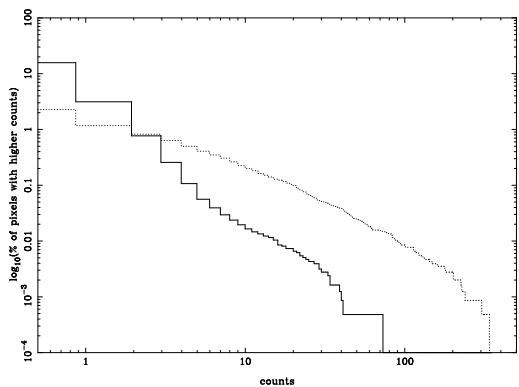}}
\caption{Percentage of the number of pixels with a count level higher than
a given value, in the weak X-ray cluster ABCG 194 (full line) and in the
bright X-ray cluster ABCG 85 normalized to the same total number of counts. }
\protect\label{histopix2}
\end{figure}

In X-rays, the signal to noise ratio is low: we give in
Fig. \ref{histopix2} the curves which allow to obtain the number of pixels
above a certain count level. 99\% of the pixels contain only 1 or 2
counts (essentially noise) for both clusters; however, there are much
more pixels with a large number of counts (essentially signal) in ABCG
85 than in ABCG 194.

\subsection{Radio data}

In the center of this cluster are located the radiosources 0123-016A
and 0123-016B. The ``dumbbell" source 0123-016A coincides with two
galaxies, one of them being NGC 547 (3C~40), while 0123-016B coincides
with NGC 541 (Ledlow \& Owen 1995, Edge \& R\"ottgering 1995); NGC 547
and NGC 541 are the first are third magnitude galaxies in the cluster
respectively.  The radiosource 3C~40 has a ``Twin Jet" structure
(Burns et al.  1994).  The optical disk of NGC 547 is perpendicular to
the radio axis (Zirbel \& Baum 1998) and its direction coincides with
that joining NGC 541 and 547 (Fasano et al.  1996).  NGC 541 is a
narrow-angle-tailed (NAT) radio galaxy.  The direction of the tail
coincides with that between NGC 541 and 547 (O'Dea \& Owen 1985).
According to Brodie et al.  (1985) and van Breugel et al.  (1985) this
radio source may be interacting with Minkowski's object.

\section{Morphological analysis}\label{morpho}

\begin{table*}[t!]
\centering
\caption{Equivalent radius, number of galaxies located in the disc (see 
text), minor to major axis ratio, direction of the major axis and center 
of the ellipse (in arcseconds relative to the cluster center). } 
\label{directionopt}
\begin{tabular}{lccc}
\hline
		& Disc$_{1}$ & Disc$_{2}$ & Disc$_{3}$  \\
		\hline
		& & &\\
		R$_{\mathrm{final}}$ (arcsec) & 370 & 1600 & 2860  \\
		Number of galaxies & 26 & 42 & 78  \\
		$b/a$ & 0.2 & 0.8 & 0.8  \\
		Direction angle & 52 & 45 & 85  \\
		Centre (X Y) & 
\begin{tabular}{c c}
			260 & -218  \\
\end{tabular}
 &\begin{tabular}{c c}
			118 & -334  \\
\end{tabular}  & \begin{tabular}{c c}
			-272 & -354  \\
\end{tabular}  \\
			& & &\\					
\hline
\end{tabular}
\end{table*}

We present below the global morphological properties of ABCG 194 both
from optical and X-ray data. After obtaining maps of the various
components, we will discuss the main cluster orientations at various
scales, and compare the density profiles for the galaxy distribution
and for the X-ray gas.

\subsection{Optical analysis}\label{morphoopt}

\begin{figure}
\centerline{\psfig{figure=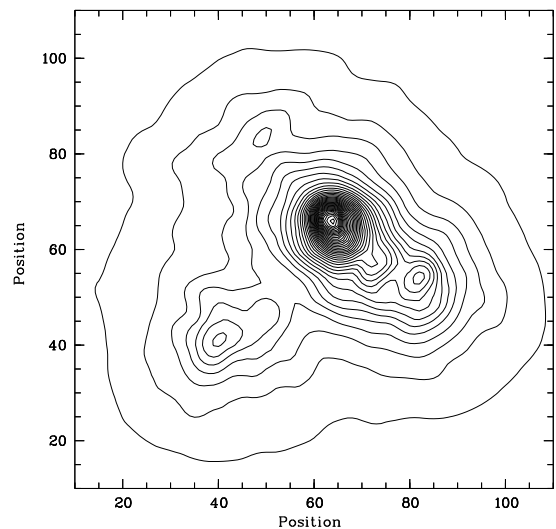,height=7cm}}
\caption{Density map of the distribution of galaxies with velocities in
the cluster obtained with a Dressler parameter of 10. The field size is that 
covered by the sample defined in section 2.1., and therefore 1~pixel
represents roughly 2 arcmin.}
\protect\label{dressler}
\end{figure}

The sample used in this subsection is the total one of 97 galaxies.

In order to illustrate the galaxy distribution, we have derived a
Dressler map of the galaxy distribution (Dressler 1980), the main
directions at various radii and the galaxy density profile.

The Dressler smoothing method has been applied to the sample of
galaxies with velocities in the cluster; the result is displayed in
Fig.~\ref{dressler}.  One can see that in the central part the cluster
is elongated along PA$\sim$50 with a second component visible in the
south east, along a direction roughly perpendicular to the main one.

We estimate these quantities with a momentum method, using the
Salvador-Sol\'e \& Sanrom\'a (1989) software: from a distribution of
points in a disc of radius $R_{\mathrm{initial}}$, this method gives
the ellipticity and major axis direction, as well as the points which
are involved in the actual ellipse (displayed in
Fig.~\ref{orientation}a), leading to an equivalent radius
$R_{\mathrm{final}} = \sqrt{a b}$.  The position angles and axial
ratios -~defined as the ratio of the small axis $b$ to the large axis
$a$~- for the three concentric discs of radii $R \leq 1800$, $R\leq
2000$ and $R\leq 3600$ arcsec are given in Table \ref{directionopt}.
The axial ratio is seen to increase with radius, showing that despite
its classification as ``linear'' ABCG 194 is elongated only in its
central parts.

\begin{figure*}[tbp]
\caption{Left panel (a): Ellipse drawn as explained in the text.  The 
corresponding equivalent radii are indicated in Table 
\ref{directionopt}.  Right panel (b): Multi-scale analysis of the 
X-ray image showing the components at the following scales: full line: 
2-pixel scale; dotted line: 4-pixel scale; dashed line: 8-pixel scale; 
dot-dashed line: 16-pixel scale; long dashed line: 32-pixel scale.  
The size of each pixel is 15$\times$15 arcseconds.  For each scale, 
two levels are drawn, corresponding to 3$\sigma$ and 12$\sigma$ of the 
background.}
\label{orientation}
\end{figure*}

\begin{table}[t]
\centering
\caption{Values for the $\beta$-model distribution of
galaxies. Notes: (1) Chapman et al.  (1988), the cluster is
supposed to be spherical; (2) $\beta$ fixed to 1, but the axial ratio
$b/a$ is free; (3) $\beta$ and the axial ratio are both free; (4) in a
rectangle of 10000$^{2}$ arcsec$^{2}$, with the $b/a$ ratio fixed at
the previous value; (5) in a rectangle of 8000$^{2}$ arcsec$^{2}$.}
\begin{tabular}{llrlc}
\hline
n$_{0}$ & ~~$\beta$ & r$_{\mathrm{c}}$~~ & $b/a$ & Notes  \\
 $(\mathrm{N/deg^{2}})$ &  & (arcsec) &  &   \\
			\hline
		 150 & 1 &    720&1 & (1)  \\
		 330 & 1 &    518 & 0.8 & (2)  \\
		 200 & 1.2 & 1050& 0.8 & (3)  \\
		 215 & 1.13 & 912 & 0.8 & (4)  \\
		 305 & 0.95 & 526 & 0.8 & (5)  \\
\hline
\end{tabular}
\label{beta}
\end{table}

Chapman et al. (1988) have fit the ABCG 194 profile (for azimuthally
averaged counts) with the following model:
	$$ \sigma(r)\ = \sigma(0)[1+(r/r_c)^2]^{-1}$$
(resulting parameters are indicated in Table~\ref{beta}).
The choice of this profile is questionable in
comparison with a more general $\beta$-model:
$$ \sigma(r)\ = \sigma(0)[1+(r/r_c)^2]^{-((3 \beta -1)/2)}.$$ However,
recent profile fits obtained from the ENACS data are consistent with a
King model with a small dispersion on the values of $\beta$ (Adami et
al.  1998). C. Adami has kindly performed several fits to our data;
his results are given in Table~\ref{beta} (elliptical profiles are
used, assuming $r = \sqrt{a b}$.  It is interesting to notice that
even when $\beta $ is free in the fitting process (rows 3-5), its
value is close to 1, in agreement with ENACS values. The axial ratio
is 0.8, in agreement with the values given in Table
\ref{directionopt}.

With $\sigma_{los} $ = 360 \kms and kT$_{\rm X}$=2.6 keV, we can
calculate $\beta _{spec}$ and find a value of 0.93, consistent with
the values given in Table \ref{directionopt}.

\subsection{X-ray analysis}\label{xana}

Keeping in mind the weakness of the signal to noise ratio, we have
tried to derive as much information as possible from the X-ray data.

\begin{figure}[tbp]
\centerline{\psfig{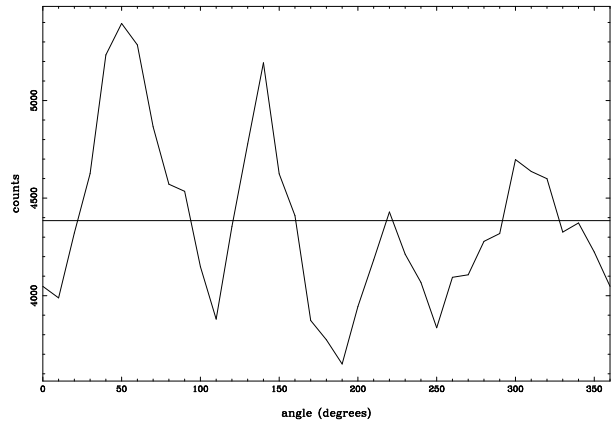}}
\caption{Sum of counts in angular sectors $30^\circ$ wide, with steps
of $10^\circ$. The center is that of the X-ray $\beta$-model, angle
$0$ is north, and angles increase counterclockwise. The image was
limited to a circle of $53$ arcmin radius. The number of counts per
sector averaged over the whole image is shown as a straight line for
comparaison.}
\protect\label{sectorx}
\end{figure}

\begin{figure}[tbp]
\centerline{\psfig{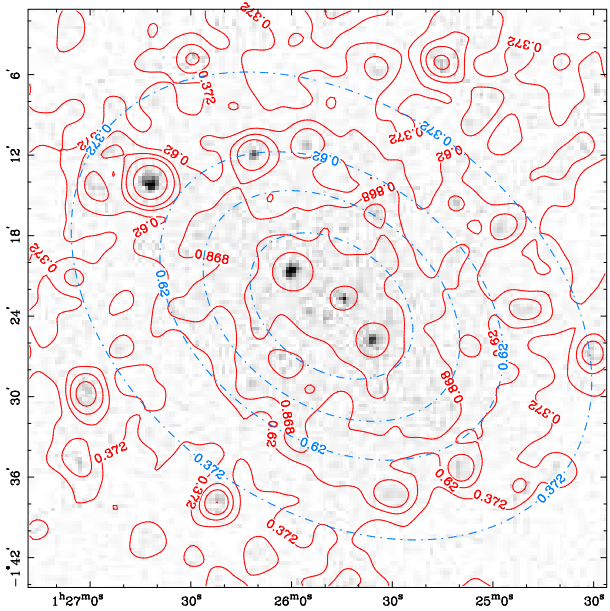}}
\caption{ROSAT PSPC image (grey scale) with isocontours of the same
image smoothed by a gaussian of FWHM $=7.56$~pixels $=113$'' (full lines).
The isocontours of the best pixel by pixel two dimensional 
$\beta$-model fit are shown as dot-dashed lines.}
\protect\label{fitx}
\end{figure}

\begin{figure}[tbp]
\centerline{\psfig{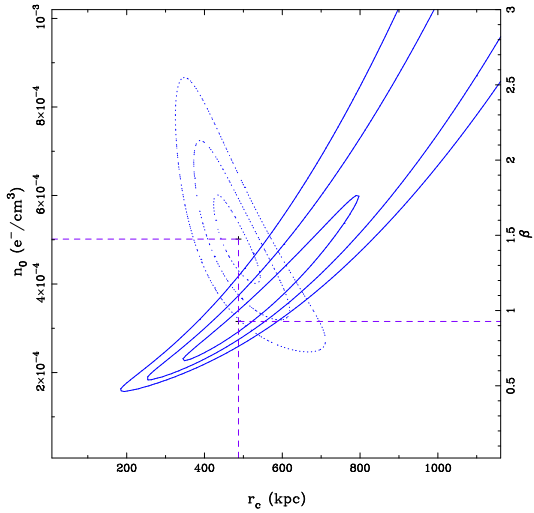}}
\caption{$1\sigma$, $2\sigma$ and $3\sigma$ $\chi^2$ contours of the 1D
$\beta$-model fit for the paramaters $r_c-n_0$ (dotted) and $r_c-\beta$ (full
line) around the values of $\chi^2_{\rm min}$, shown by the dashed lines.
}
\protect\label{cont_chi2}
\end{figure}

While two privileged directions are found from the optical analysis
(see previous subsection), only one is found at first sight from the
X-ray image. We have therefore made an analysis of the number of
counts in angular sectors centered on the X-ray $\beta$-model center
and $30^\circ$ wide, rotating with a step of
$10^\circ$. Fig. \ref{sectorx} shows the existence of four peaks; the
two strongest peaks correspond to PA=50 (the main principal direction
of the cluster) and PA=140 pointing towards the south east optical
enhancement. The third peak, which is symmetrical to that at PA=50 is
very weak. The fourth peak is symmetrical to the second one, and shows
that there is also X-ray emission towards the north west.  Note that
the first peak is certainly contaminated by the bright X-ray source
labeled F in Fig. \ref{orientation}b.

We performed a multi scale wavelet analysis on this image to eliminate
the noise and identify structures at various scales (see a full
description of the method e.g. in Slezak et al. 1994).  This was done
in the central 35' diameter region (within the PSPC annulus), as
illustrated in Fig.~\ref{orientation}b.  This obviously corresponds
only to the central part of the cluster (see Fig.~\ref{xya194}).  The
limiting radius is $R_{\rm L}\simeq$1540~arcsec, or 0.8 Mpc.

At the largest scale (32 PSPC pixels, or 4 arcmin), one can observe 
elliptical emission with a major axis of about 50$^\circ$ and
an axial ratio $b/a \simeq 0.66$.

At a smaller scale (16 PSPC pixels, or 2 arcmin), the general behavior
is similar. Due to to the weight of the powerful X-ray emitting
point-like source (F), the (X,Y) origin has shifted toward the west
along the major axis.

\begin{table}[tbp]
\centering
\caption{Results of different fits of the diffuse emission over the whole
image. The sources detected by Snowden's software were not included in these
fits. }
\begin{tabular}{lccccc}
\hline
 Model        &  n$_{0}$&$r_\mathrm{c}$& $\beta$  & $b/a$      & PA     \\
              & ($10^{-4}e^-/cm^3$)& (kpc)       &        &       & ($^{\circ}$)\\
\hline
              &         &             &              &            & \\
$\beta$ model & 5       & 524         &  1.02        &  0.7       & 54    \\
King          & 5       & 515         &  1           &  0.7       & 54    \\
\hline
              &         &             &              &            & \\
1D $\beta$ model& $5$ & $480$ & $0.90$ &            & \\
$1\sigma$ errors&$^{+1.45}_{-1.1}$& $^{+646}_{-185}$  & $^{+1.84}_{-0.27}$   &            & \\
\hline
\end{tabular}
\label{betaX}
\end{table}

The temperature was fixed to the constant value of 2.6$\pm$0.15 keV
measured by ASCA (Fukazawa et al. 1998). We calculated the bolometric
X-ray luminosity by normalizing a photo-electric absorbed
(ga\-lac\-tic $n_{\rm H}=3.77\ 10^{20}{\rm cm}^{-2}$) {\tt mekal}
model (metal abundances~= 0.3Z$_\odot$, $n_e=5\ 10^{-4} {\rm
cm}^{-3}$) to the source masked and background subtracted total
diffuse emission (found to be $0.42$ counts/sec).  We find
$(1.5\pm0.1) 10^{43}$erg s$^{-1}$.

We have performed a pixel by pixel fit to the data as described by
Pislar et al. (1997). A $\beta$-model was used for the density law,
including or not the strong point-like sources. The sources detected
by Snowden's software were not included in these fits. The center is
found at coordinates: $\alpha _{2000}=1^h25^{mn}49^s$, $\delta
_{2000}= -1^\circ 23'30''$, i.e. displaced to the south west
relatively to the optical center given by CGH. The background found
($\sim$0.24 counts/pixel) is consistent with the mean background taken
in an annulus far enough from the center to be flat on a wide
area. Results are shown in Fig. \ref{fitx}: the isocontours appear
rather satisfactory, but the various parameters are not well
constrained (the {\tt MINOS} routine from {\tt MINUIT} was unable to
find error boundaries).  We then made a $\beta$-model fit on the
profile obtained by summing the counts in elliptical bins (with their
center, PA and ellipticity given by the pixel by pixel fit).  This
$\chi^2$ fit leads to similar values, and to very large $1\sigma$
error bars. Fig. \ref{cont_chi2} shows how badly $r_c$ is constrained
: the $2\sigma$ contour of $\chi^2(r_c,\beta)$ stays open for very
large values of $r_c$.  The $1\sigma$, $2\sigma$ and $3\sigma$ values
are defined for $\chi^2_{\rm red_{\rm min}}+1$, 4 and 9 respectively.

Note that the ellipticity found in X-rays is comparable to that in
the optical (see Table~\ref{betaX}). 

\subsection{Masses} 

\begin{figure}
\centerline{
\psfig{figure=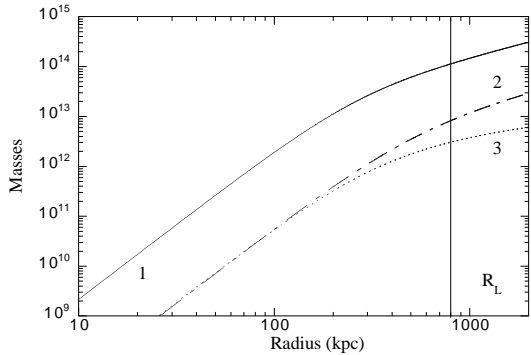,width=7cm}} 
\caption{Mass profiles. 1 is the dynamical mass profile
assuming that the X-ray gas is isothermal; 2 - X-ray gas mass; 3 -
stellar mass. Error bars on these quantities at the R$_{\rm L}$ radius
are given in the text. }
\protect\label{lesmasses}
\end{figure}

The X-ray gas mass M$_{gas}$ was calculated by integrating the
$\beta$-model given in Table 3. At the limiting radius R$_{\rm L}$=0.8
Mpc (defined as the radius where the X-ray background is reached), the
X-ray gas mass is M$_{gas} = 9\ 10^{12} h_{50}^{-5/2}$ M$_\odot$, 
with limits correponding to 1$\sigma$ errors:
$6\ 10^{12} \leq $ M$_{\rm gas} \leq 2\ 10^{13} \ h_{50}^{-5/2}$ M$_\odot$.
The very large error bars are mostly due to the fact that the $r_c$
parameter is not well constrained.

The stellar mass M$_{stellar}$ was calculated by integrating the King
model given in Table 2, assuming M/L$_{\rm B}$=10 M$_\odot$/L$_\odot$.
At the X-ray limiting radius R$_{\rm L}$, the stellar mass is $(3.0
\pm 0.8)\ 10^{12}\ h_{50}^{-2}$ M$_\odot$. However, the galaxies
extend much further out, up to a radius of about 3 Mpc, where the
corresponding stellar mass is $(7.5 \pm 1.9) 10^{12} h_{50}^{-2}$
M$_\odot$.  Note that these values are not well constrained: besides
the usual errors on the central density and core radius, the dominant
source of error is the incompleteness of our data. In the central part
of the cluster (for radii smaller than 2100 arcsec) faint galaxies are
relatively more numerous than in the outskirts: 9 galaxies have
magnitudes fainter than 16 while 34 are brighter than this value.  The
stellar luminosity inside radius R is: ${\rm L(<R) \propto n_0 R_c^3
I}$, where I is the integration of the King distribution.  The error
on the luminosity is in fact that on $n_0 R_c^3$, which is independent
of radius; the error on I is negligible. We therefore consider that
the error on the stellar mass due to incompleteness at faint
magnitudes is at most 25\%.

The M$_{gas}$/M$_{stellar}$ ratio is about unity for radii smaller
than $\sim$200 kpc, and reaches a value of $\sim 3\pm 0.3$ for R$_{\rm
L}$ (applying error propagation equations). 

The hydrostatic equation applied to the X-ray gas allows us to
estimate the total cluster dynamical mass, with hypotheses on the
X-ray gas temperature. Since we have no X-ray gas temperature map, we
will assume the gas to be isothermal. The corresponding mass profile
is displayed in Fig. \ref{lesmasses}. The dynamical mass at radius
R$_{\rm L}$ is M$_{dyn} =8\ 10^{13}\ h_{50}^{-1}$M$_\odot$,
with limits correponding to 1$\sigma$ errors:
$3\ 10^{13} \leq $ M$_{dyn} \leq 2\ 10^{14} \ h_{50}^{-1}$ M$_\odot$.  
Although the dynamical mass derived from the X-rays is not very
accurate, it agrees with the virial mass estimated by Girardi et al. (1998).

The ratios of the X-ray and stellar masses and of the sum of these two
quantities to the dynamical mass are 0.12, 0.04 and 0.16 respectively
(for $h_{50}=1$) at radius R$_{\rm L}$. These values are only
indicative within a factor of 2, due to the large error bars on the
masses.

\section{Structural properties}

The presence of substructure in the ABCG 194 cluster of galaxies has
been a subject of debate:

- Using an adaptive kernel algorithm, Kriessler \& Beers (1997) 
determined the presence of substructure in the central core of this 
cluster with a probability level of 95\%.

- By applying a multiscale analysis which couples kinematic estimators 
with wavelet transforms (Escalera et al. 1994), Girardi et al.  (1997) 
classify the ABCG 194 cluster as unimodal with a very condensed core 
and two small-scale subgroups within the cluster.

- Beers \& Tonry (1986) have suggested that this cluster shows good
evidence for multiple X-ray substructure.

At a smaller scale -~thus not really corresponding to substructures~-
from a 40'$\times$40' Digital Sky Survey image around the ABCG 194
cluster center and applying a wavelet analysis technique to the
corresponding ROSAT PSPC image, Lazzati et al.  (1998) detected 26
X-ray sources; four of these coincide with known galaxies in the
cluster: NGC 547, NGC 541, NGC 564 and NGC 538.

\subsection{Optical structure} \label{structopt}

An overall description of the Dressler map has been given in section
\ref{morphoopt}.  Besides the large scale morphological aspect
discussed above, in particular the bright extension along PA=50 which
accounts for the ``linear'' aspect of the cluster, we also see a
fainter extension towards the south-east.

This tends to indicate the presence of substructure. In order to
qualify the degree of significance of this substructure, we have
applied the spatial test developed by Salvador-Sol\'e et al. (1993),
which has been shown to be well suited to the detection of
substructure in systems (Scodeggio et al. 1995).  By applying this
method to the total sample of 97 galaxies, we estimate two density
profiles: the first one, N(r)$_{\rm dec}$, is obtained by inverting
the density of numbers of pairs of galaxies, and the second one,
N(r)$_{\rm dir}$, by inverting the density of numbers of galaxies at
projected distances from the center of symmetry.  N(r)$_{\rm dir}$ is
therefore insensitive to correlations in galaxy positions. A
difference among these two quantities is interpreted as indicating the
presence of substructure in the system.

\begin{figure}[tbp]
\centerline{
\psfig{figure=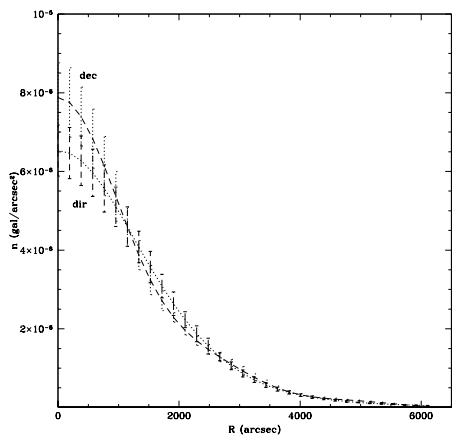,width=6cm}}
\caption{N(r)$_{\rm dec}$ and N(r)$_{\rm dir}$ profiles (see section 3.1). }
\protect\label{perfild}
\end{figure}

The distributions of N(r)$_{\rm dec}$ and N(r)$_{\rm dir}$ are
displayed in Fig. \ref{perfild}; they are indistinguishable within
error bars (estimated with a Monte Carlo technique). We have applied
the statistical method proposed by Salvador-Sol\'e et al. (1993) to
test the significance of substructures at a scale chosen \textsl{a
priori}. We find that already at a scale of 0.25 Mpc the probability
to have substructure is less than 30\%, and this probability strongly
decreases with increasing scale. Therefore this method detects no
significant substructures in the optical sample (see Salvador-Sol\'e
et al. 1993 for details). Note that the result is the same if we apply
this method to a more complete magnitude limited subsample.

\subsection{Substructures derived from kinematics}

\begin{figure*}[t!]
\centerline{\psfig{figure=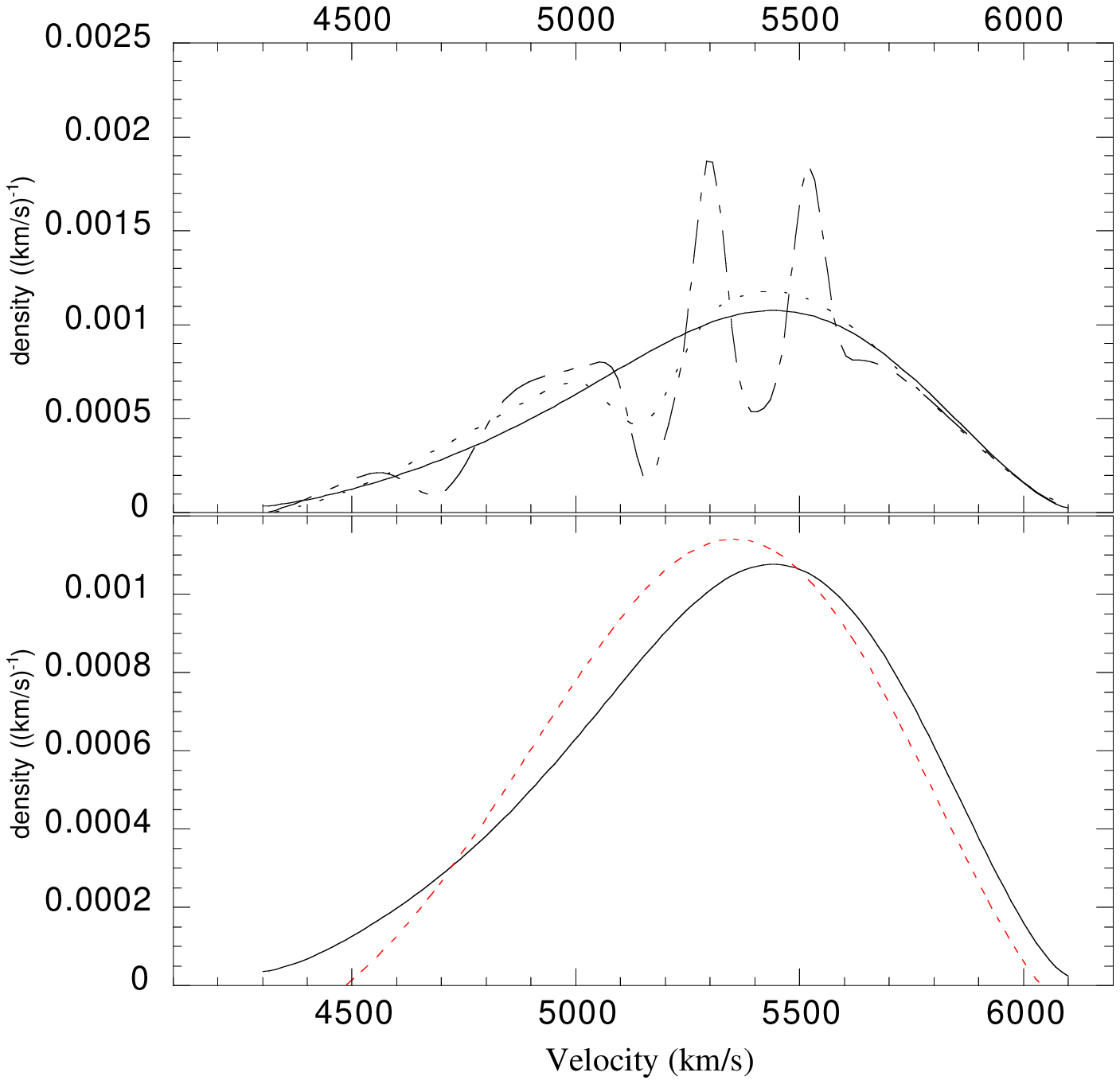,height=6cm} \qquad
\psfig{figure=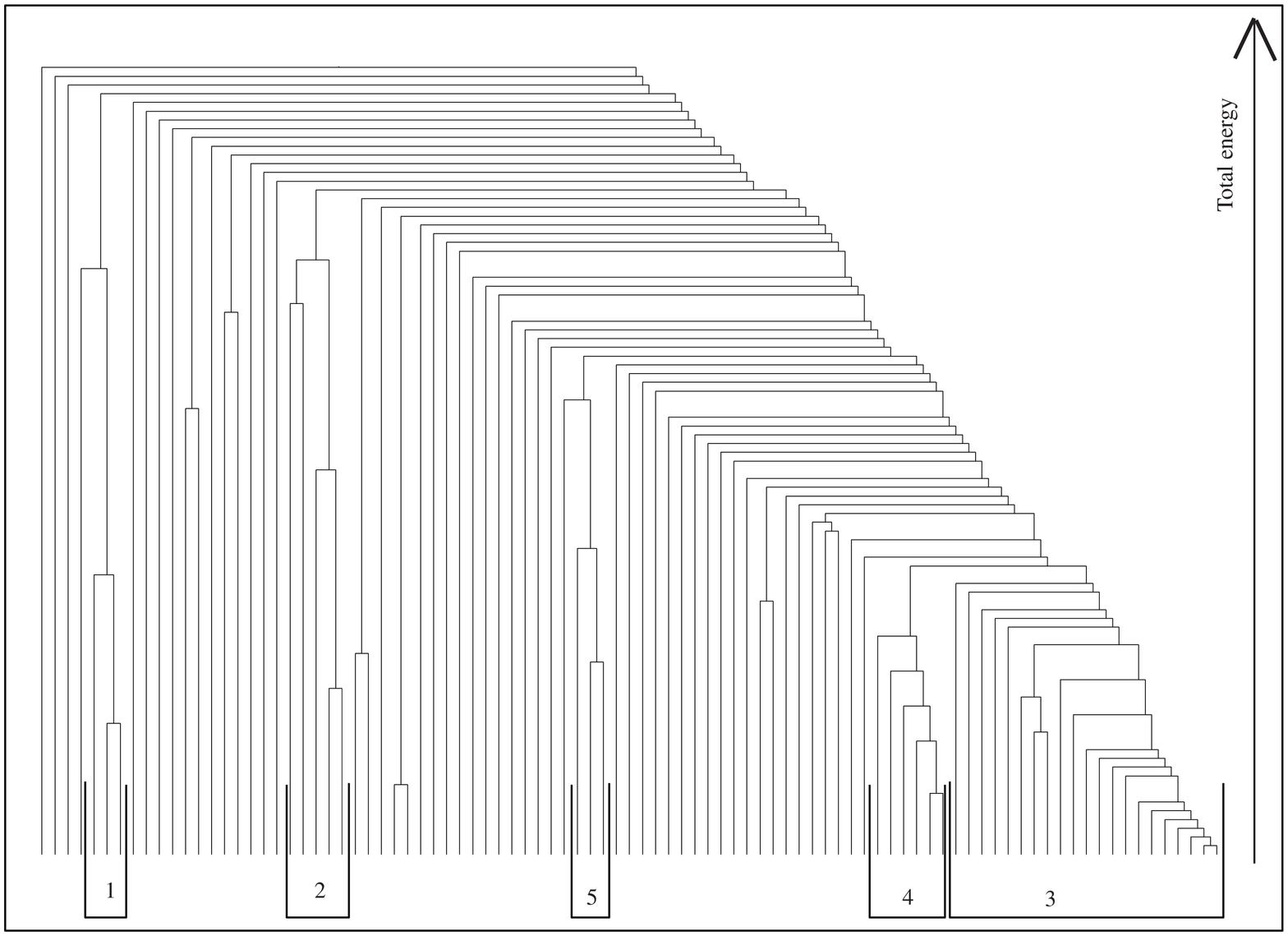,height=8cm,angle=90}}
\caption{Left: velocity density distribution of galaxies. Top panel (a):
galaxies belonging to the ``All'' sample (see text); the three
densities correspond to confidence levels: 3$\sigma$ (full line),
2.6$\sigma$ (dotted line) and 2.2$\sigma$ (dot-dashed line).  Bottom
panel (b): galaxies belonging to the All (full line) and Main (dashed
line) samples at a 3$\sigma$ level. (c) Right: dendogram obtained by the HTree 
method for 91 galaxies. The numbers at the bottom correspond to those of the 
groups defined in Table \ref{groupes}.} 
\protect\label{wave1a194}
\end{figure*}

The previous analysis considers the cluster globally but does not take
into account the dynamical properties of substructures (if any). We
will therefore apply the Serna \& Gerbal (1996) method, coupled with a
study of the galaxy velocity distributions in the cluster.

This method sorts the galaxies according to their total energy
(i.e. the sum of the potential and kinetic energies), leading to a
dendogram where the total energy appears vertically.  Pairs and groups
of galaxies then appear with a lower total energy.  The velocity
density distributions were obtained using profile reconstructions
based on a wavelet technique (instead of a histogram). The features
obtained with this method are significant at various chosen levels
above the noise (Fadda et al. 1998).

We applied this method to a sample of 97 galaxies defined as follows:
the CGH velocity sample of 74 galaxies with both velocities in the
cluster range and magnitudes; the BCG sample of 22 galaxies for which
magnitudes were not available except for two which we found in the NED
database; for the remaining 20 we assigned each galaxy the mean
magnitude of the CGH sample: 15.2; one Seyfert galaxy (Knezek \&
Bregman 1998) to which we also assigned a magnitude of 15.2.

A first dendogram reveals the presence of six galaxies which appear to
be only very loosely linked to the cluster.  The velocities of these
six galaxies are quite similar, but they are distant from each other
in projection on the sky.  The velocity distribution indicates a bump
in the 6500 km/s region, well separated from the general distribution.
We therefore chose to eliminate these objects in the following
analysis, and are left with a sample of 91 galaxies; this sample leads
to the velocity interval 4300-6200 km/s for galaxies belonging to the
cluster, that is a somewhat narrower range than found by CGH.  This
sample will be referred to as the ``All'' sample, hereafter.

The velocity distribution of this new sample at various significance
levels is displayed in Fig.~\ref{wave1a194}a.  The overall shape of
this distribution is obviously not gaussian, but shows an asymmetry
with an excess at high velocities. This is confirmed by the values of
the skewness and kurtosis: see values for ``All'' in Table~4.  Its
mean velocity (which is also the median) is equal to that of the
second brightest galaxy (NGC 541).

The dendogram obtained after excluding these six galaxies (see Fig.
\ref{wave1a194}c) reveals the presence of subgroups.  Groups 1 to 5
(in order of increasing mean velocity) come out of the sample
easily. We have constructed a ``Main'' structure by taking out these
five groups from the ``All'' sample.  The statistical characteristics
for these various subgroups are presented in Table~4 and a map of
group positions in the overall field is shown in Fig.~13.

\begin{table*}
\begin{center}
\caption{Characteristics of the subgroups identified from the
dendogram in Fig. \ref{wave1a194}c. Columns have the following
meaning: (1) group name, (2) number of members in group, (3) average
radial velocity of group in km/s, (4) velocity standard deviation in
km/s, (5) skewness, (6) kurtosis.}
\begin{tabular}{lrrrrr}
\hline
 Name      &   Nb.   &$<$Vr$>$& St.Dev.& Skewness & Kurtosis \\
           & of gal. & (km/s)& (km/s)  &          & \\
           &         &       &         &          & \\
\hline
           &         &       &         &          & \\
All        &   91    & 5326  &  361    &  -0.37   & -0.48 \\
Group 1    &    5    & 4643  &  189    &          & \\
Group 2    &    3    & 4805  &  166    &          & \\
Group 3    &   21    & 5343  &  304    &   -0.07  & -0.01 \\
Group 4    &    6    & 5655  &  145    &          & \\
Group 5    &    3    & 5895  &   54    &          & \\
Main       &   53    & 5330  &  303    &  -0.23   & -0.53 \\
\hline
\end{tabular}
\end{center}
\label{groupes}
\end{table*}
Group 3 appears to be the central subsystem, at the bottom of the
gravitational potential well of the cluster. It contains the three
brightest galaxies (including NGC 541 and NGC 547 which are both
radiosources) which show a high level of boundness. Notice the
presence of three other galaxies which appear to be bound together,
among which the X-ray galaxy NGC 538, and also that of the Seyfert
galaxy.

Groups 1 and 4 are well defined both spatially and in velocity space
(their velocity dispersion is small). Group 4 contains the X-ray
galaxy NGC 564.

Groups 2 and 5 contain only a few galaxies with small radial velocity
dispersions and average intergalactic distances.  Note that group 2 is
far from the cluster center and appears to be weakly bound to the bulk
of the cluster, as seen on the dendogram.

In comparison with the ``All'' sample, the ``Main'' structure velocity
density distribution appears more gaussian, with a mean velocity equal
to that of the overall sample within the error bars, but a smaller
velocity dispersion and skewness (see Fig. \ref{wave1a194}b). The
center of the ``Main'' sample is displaced towards the north east
relatively to the ``All'' sample.

Our results can be compared to those found in the literature and based
on various methods. Using a technique combining wavelets and
kinematical information, Girardi et al. (1997) determined the
small-scale structure in the ABCG 194 cluster and found the same group
as our group 2, and a group which, by coordinates and radial velocity
coincides with our group 3. Using two other different methods to
extract groups, Garcia (1993) also found our group~3. This group
therefore appears to be at the bottom of the cluster potential well,
while the other groups described in Table~4 appear more like pairs of
galaxies with possible satellites.

Notice that the ``linear'' structure of the cluster corresponds to our
group 3, while the extension observed towards the south east in the
Dressler map (Fig. \ref{dressler}) may correspond to our group~4.

The velocity dispersion profile (VDP) is shown in
Fig. \ref{dispersion} for the ``All'' sample. The binning has been
performed in ellipses with their major axis along the direction PA=50
and with an axial ratio $b/a = 0.8$; the result is shown for bins of
equivalent radii 400 and 800 arcsec. The shape of this profile does
not change (within the error bars) if the subclusters are excluded.
The VDP ``inverted'' shape shows an increase with radius up to
$\approx 2000$ arcsec, then a decrease and is comparable to that
derived by den Hartog \& Katgert (1996) for this cluster.  These
authors interpret such a profile as originating from a relaxed region.

\begin{figure}[tbp]
\vskip 1.6truecm
\caption{Positions of subgroups in the overall cluster field.
The numbers of the groups indicated correspond to those in Table 
\ref{groupes}. }
\protect\label{groupsa194}
\end{figure}

\begin{figure}[tbp]
\centerline{\psfig{figure=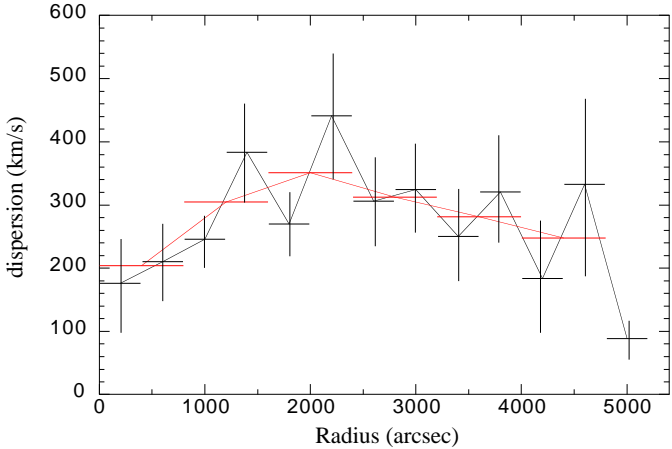,height=6cm}}
\caption{Velocity dispersion as a function of radial distance to the
cluster center; the equivalent radius of each bin is equal to 400 arcsec for
the bold curve and to 800 arcsec for the light one. }
\protect\label{dispersion}
\end{figure}

\subsection{X-ray structure}

We now compare these results with the X-ray features derived first
from the wavelet analysis and from the ``detect'' software developed
by Snowden, and second from the pixel by pixel fit.

\begin{table}
\begin{center}
\caption{Luminosity of the sources detected in the hard band
(0.44-2.04 keV).  The X-ray luminosity is computed with a 1keV mekal
model. }
\begin{tabular}{cccl@{}ll@{}l}
\hline Source & $\alpha _{2000}$ & $\delta _{2000}$ & \multicolumn{2}{c}{counts/ksec} & \multicolumn{2}{c}{$L_X$} \\
              &                  &                  &     & &\multicolumn{2}{c}{($10^{41}$erg/s)}\\
\hline
 A & $1^h26^{mn}00^s$ & $-1^\circ20'55''$ & 16.8~ &$\pm1.1$ & 3.1 &$\pm0.2$\\
 B & $1^h25^{mn}45^s$ & $-1^\circ22'56''$ & 5.6  &$\pm0.8$ & 1.0 &$\pm0.1$\\
 C & $1^h25^{mn}37^s$ & $-1^\circ26'09''$ & 9.6  &$\pm0.9$ & 1.8 &$\pm0.2$\\
 D & $1^h26^{mn}11^s$ & $-1^\circ12'07''$ & 5.8  &$\pm0.8$ & 1.0 &$\pm0.1$\\
 E & $1^h25^{mn}57^s$ & $-1^\circ11'18''$ & 2.3  &$\pm0.5$ & 0.4 &$\pm0.1$\\
 F & $1^h26^{mn}43^s$ & $-1^\circ14'15''$ & 21.5 &$\pm1.2$ & 4.0 &$\pm0.2$\\
 G & $1^h26^{mn}23^s$ & $-1^\circ38'11''$ & 3.8  &$\pm0.6$ & 0.7 &$\pm0.1$\\
 H & $1^h25^{mn}31^s$ & $-1^\circ37'29''$ & 3.5  &$\pm0.5$ & 0.6 &$\pm0.1$\\
 I & $1^h25^{mn}10^s$ & $-1^\circ35'37''$ & 2.6  &$\pm0.6$ & 0.5 &$\pm0.1$\\
\hline
\end{tabular}
\end{center}
\label{allx}
\end{table}

The wavelet analysis of the X-ray image reveals no substructure at
middle and large scales.  At the two smallest scales (2 and 4 pixels,
or 30 and 60 arcsec), 9 components (at least) are detected at a
3$\sigma$ level (see Fig. \ref{orientation}b). The Snowden method
gives 38 sources in the same field, among which 9 in common with
ours. The difference between the numbers of detections reached by both
methods raises the question of their validity.  The Snowden method is
based on a convolution by the PSF (which varies with radius), and is
therefore close to the wavelet method. However, the statistics on
which the detection thresholds are determined for both methods are not
the same. Our wavelet software estimates the image characteristic
statistics by analyzing the whole image (see e.g. Slezak et al. 1994)
while Snowden considers only circles containing between 90\% and
$2.5\times90\%$ of the encircled energy radius of the off-axis
PSF. The average background is taken into account only in the cases
where there are less than 4 counts per annulus.  This may explain the
difference between the numbers of sources. Note also that the purpose
of both methods is not the same: Snowden's software is aimed at
detecting everything that is not diffuse X-ray emission, in order to
eliminate these point-like sources, and therefore it tends to find a
higher number of small sources (including possible cosmic rays), while
our purpose is to detect only X-ray sources above a certain
significance level.

Source A corresponds to one of the two ``dumbbell'' galaxies, which
are the first and third brightest cluster galaxies; however the X-ray
PSPC pixel size is too large to be able to discriminate between both
galaxies.  Source B can be identified with the second brightest galaxy
in the cluster. Source C coincides with the Seyfert galaxy reported by
Knezek \& Bregman (1998).  The radiosources 0123-016AB are also strong
X-ray emitters (Burns et al. 1994).  Note that these sources have
velocities in the cluster. Source F appears to be a star (it is
star-like in the POSS and there is no quasar at this position in the
V\'eron-Cetty \& V\'eron 1993 catalogue).  The other sources do not
coincide with any galaxies with available redshifts.  Note that in the
part of the X-ray field shown in Fig.~\ref{orientation}b, we detect
all the sources reported by Lazzatti et al. (1998) except one (their
source 26).

We have estimated the X-ray luminosity of the nine detected sources,
proceeding in three steps:
\begin{itemize}
\item we fit the whole data by a $\beta$-model as explained in section 3.2;

\item we subtracted  to the real image the synthetic image corresponding to 
the values obtained by the fit;

\item in the image thus obtained, we have counted the photons around
the peaks corresponding to the detected sources and give the
corresponding results in Table~5; we have checked that outside the
sources the background is around zero, as expected;

\item the luminosities were derived from the counts assuming a
temperature of 1 keV, following the method described in section
\ref{xana}.

\end{itemize}

The luminosities found for these sources are much lower than those
given by Lazzati et al. (1998), but the reason for such a discrepancy
is not clear. We have compared our source counts with the results
given by the ROSAT SASS data processing pipeline, and find that they
agree within 35\% except for sources B and C. Note that the specific
$\beta$-model used for this estimate is not very important.

\section{Conclusions}

We have presented a morphological and structural analysis of ABCG 194
both at optical and X-ray wavelengths. From dynamical and velocity
dispersion studies, we are left with a sample of 91 galaxies really
belonging to the cluster (4200-6100 km/s), the mean cluster velocity
being 5326 km/s with a standard deviation of 360 km/s. This value
coupled with the temperature of 2.6 keV determined for this cluster
from ASCA data falls on the $\sigma$-T$_{\rm X}$ relation (Wu et
al. 1998). The main underlying structure (53 galaxies) has the same
mean velocity but a standard deviation of 300 km/s. The X-ray map
shows an ellipticity similar to that observed in the optical.

The X-ray luminosity of ABCG 194 is $1.5\ 10^{43}$erg s$^{-1}$,
and falls notably below the L$_{\mathrm{X}}$-T$_{\mathrm{X}}$
relation for clusters (Markevitch et al. 1998, Arnaud \& Evrard 1999),
i.e. this cluster is underluminous for its temperature. On the other
hand, the cluster is within the dispersion of the
L$_{\mathrm{X}}$-$\sigma$ relation (Mahdavi et al. 1997). The
$\beta$-model fit to the X-ray gas gives $\beta =1.02$, which appears
larger than expected from the low value of the X-ray gas temperature
(Arnaud \& Evrard 1999). A $\beta$-model fit to the galaxy density
distribution also gives values of $\beta$ consistent with 1, as in the
ENACS sample (Adami et al. 1998).

ABCG 194 appears therefore to be a very poor cluster, comparable to
those recently studied by Mahdavi et al. (1999), as confirmed by the
stellar, X-ray gas and dynamical masses which we have calculated.

The number of counts in angular sectors from the X-ray $\beta$-model
center (Fig. \ref{sectorx}) shows priviledged directions along the
major axis PA found in X-ray and optical analysis, but also along an
axis perpendicular to this one.  Except for individual sources, no
X-ray substructures are found with a wavelet analysis. Only 9 X-ray
sources are found with our method, while 38 are detected with the
Snowden software.

The various analyses described above show that at large scale ABCG 194
is overall a relaxed cluster, with a few superimposed groups: the
velocity distribution of the ``Main'' relaxed underlying cluster is
not far from gaussian. However, both at optical and X-ray wavelengths,
a bright linear structure is observed in the central part, and
emission is much stronger south east of this line than north west.
This central structure is well defined dynamically and is also rich in
radio and X-ray galaxies.

We can try to interpret these results in the framework of hierarchical
structure formation. As suggested by numerical simulations such as
those by Tormen et al. (1997), the time interval between two possible
major mergings is large compared to the timescale of violent
relaxation. The cluster therefore has time to relax between two
mergers. This is also the interpretation proposed for detailed
features observed for example in the Coma cluster (Biviano et
al. 1996).  We suggest that in ABCG 194 a group around the bright
elliptical galaxy NGC 547 may have merged into the cluster; due to
dynamical friction, this group has stopped near the cluster center and
was disrupted (Gonz\'alez- Casado et al. 1994). The galaxy excess
towards the south east could then be a remnant of this group.

ABCG 194 therefore appears as one of the very few poor clusters which
are well studied at various wavelengths. Such objects are intermediate
cases between rich clusters and groups, and as such are interesting to
compare the physical processes taking place in these objects, 
which may be influenced by the very different relaxation timescales of
these various systems.

\begin{acknowledgements}
We are very grateful to the Franco-Armenian PICS and Jumelage for
making our collaboration possible through financial support for three
stays of E.N.  in France.  We gratefully thank Dario Fadda and Eric
Slezak for giving us their wavelet analysis programmes, and Christophe
Adami for performing a $\beta$-model fit to our data. We also
appreciated a discussion on radio sources with Jacques Roland. Last
but not least, we thank the referee, Andrea Biviano, for several
interesting criticisms and suggestions.
\end{acknowledgements}


\begin{thebibliography}{}
\bibitem{} Adami C., Mazure A., Katgert P., Biviano A., 1998, A\&A 336, 63
\bibitem{} Arnaud M., Evrard A.E., 1999, MNRAS 305, 631
\bibitem{} Baier F.W., 1977, Astr. Nachr. 298, 151
\bibitem{} Barton E.J., De Calvalho R.R., Geller M.J., 1998, AJ 116, 1573
\bibitem{} Beers T.C., Tonry J.L., 1986, ApJ 300, 557
\bibitem{} Bird C., 1994, AJ 107, 1637
\bibitem{} Biviano  A., Durret F., Gerbal D., Le F\`evre O., Lobo C.,
et al., 1996, A\&A 311, 95
\bibitem{} Brodie J.P., Browyer S., McCarthy P., 1985, A\&A 259,L31
\bibitem{} Burns J.O., Rhee G., Owen F.N., Pinkney J., 1994, ApJ 423, 94
\bibitem{} Cavaliere A., Menci N., Tozzi P., 1998, ApJ 501, 493
\bibitem{} Chapman G.N.F., Geller M.J., Huchra J.P., 1988, AJ 95, 999
\bibitem{} David L.P., Jones C., Forman W., 1996, ApJ 473, 692
\bibitem{} Dressler A., 1980, ApJ 236, 351
\bibitem{} Durret F., Forman W., Gerbal D., Jones C., Vikhlinin A., 1998,
A\&A 335, 41
\bibitem{} Edge A.C., R\"ottgering H., 1995, MNRAS 277, 1580
\bibitem{} Escalera E., Biviano A., Girardi M. et al., 1994, Ap J 423, 539
\bibitem{} Fadda D., Slezak E., Bijaoui A., 1998,  A\&AS 127, 335
\bibitem{} Fasano G., Falomo R., Scarpa R., 1996, MNRAS 282, 40
\bibitem{} Fukazawa Y., Makishima K., Tamura T., et al. 1998, PASJ 50, 187
\bibitem{} Garcia A.M., 1993, A\&AS 100, 47
\bibitem{} Girardi M., Escalera E., Fadda D. et al., 1997, ApJ 482, 41
\bibitem{} Girardi M., Giuricin G., Mardirossian F., Mezzetti M., Boschin W.,
1998, ApJ 505, 74
\bibitem{} Gonz\'alez-Casado G., Mamon G.A., Salvador-Sol\'e E., 1994, 
ApJ 433, L61
\bibitem{} Grebenev S.A., Forman W., Jones C., Murray S., 1995, ApJ 445, 607
\bibitem{} den Hartog R., Katgert P., 1996, MNRAS 279, 349
\bibitem{} Kriessler J.R., Beers T.S., 1997, AJ 113, 80
\bibitem{} Knezek P.M., Bregman J.N., 1998, AJ 115, 1737
\bibitem{} Lazzati D., Campana S., Rosati P., Chincarini G., Giacconi R., 1998,
A\&A 331, 41
\bibitem{} Ledlow M.J., Owen F.N., 1995, AJ 109, 853
\bibitem{} Maccagni B., Garilli M., Tarenghi M., 1995, AJ 109, 465
\bibitem{} Mahdavi A., B\"ohringer H., Geller M.J., Ramella M., 1997,
ApJ 483, 68
\bibitem{} Mahdavi A., Geller M.J., B\"ohringer H., Kurtz M.J., Ramella M., 
1999, ApJ 518, 69
\bibitem{} Markevitch M., 1998, ApJ 504, 27
\bibitem{} Markevitch M., Forman W.R., Sarazin C.L., Vikhlinin A., 1998,
ApJ 503, 77
\bibitem{} Mohr J.J., Fabricant D.F., Geller M.J., 1993, ApJ 413, 492
\bibitem{} O'Dea C.P., Owen F.N., 1985, AJ 90, 927
\bibitem{Pislar} Pislar V., Durret F., Gerbal D., Lima Neto G.B., Slezak E., 
1997, A\&A 322, 53
\bibitem{} Salvador-Sol\'e E., Sanrom\'a M., 1989, ApJ 345, 660
\bibitem{} Salvador-Sol\'e E., Sanrom\'a M., Gonz\'alez-Casado G., 1993,
ApJ 402, 398
\bibitem{} Salvador-Sol\'e E., Solanes, J. M., Manrique, A., 1998, ApJ
499, 542
\bibitem{} Scodeggio M., Solanes J.M., Giovanelli R., Haynes M.P., 1995, 
ApJ 444, 41
\bibitem{} Serna A., Gerbal D., 1996, A\&A 309, 65
\bibitem{} Slezak E., Durret F., Gerbal D., 1994, AJ 108, 1996
\bibitem {} Snowden S.L., McCammon D., Burrows D.N., Mendenhall 
J.A., 1994, ApJ 424, 714
\bibitem{} Struble M.F., Rood H.J., 1982, AJ 87, 7 
\bibitem{} Tormen G., Bouchet F.R., White S.D.M., 1997, MNRAS 286, 865
\bibitem{} van Breugel W., Filippenko A.V., Heckman T., Miley G., 1985, 
ApJ 293, 83
\bibitem{} V\'eron-Cetty M.-P., V\'eron P., 1993, {\sl A catalogue of 
quasars and active nuclei}, 6th edition, ESO Scientific Report No. 13
\bibitem{} West M.J., 1994, in ``Clusters of Galaxies'', Eds. F. Durret, 
A. Mazure \& Tran Than Van, Editions Fronti\`eres
\bibitem{} West M.J., Jones C., Forman W., 1995, ApJ 451, L5
\bibitem{} Wu X.-P., Fang L.-Z., Xu W., 1998, A\&A 338, 813
\bibitem{} Zabludoff A.J., Zaritsky D., 1995, Ap J 447, L2
\bibitem{} Zirbel E.L., Baum S.A., 1998, ApJS 114,177.
\end{thebibliography}
\end{document}